\documentclass[onecolumn,11pt,draftcls]{IEEEtran}
\usepackage{amsmath,amssymb,amsthm,graphicx,graphics,subfigure,enumerate}
\usepackage{parskip}		%% blank lines between paragraphs, no indent
\usepackage{bm}
\usepackage{color}
\usepackage{epstopdf}
\usepackage{blkarray}
\usepackage{multirow}
\usepackage{mathtools}
\usepackage{url}  % Formatting web addresses
\usepackage{ifthen}  % Conditional
\usepackage{multicol}   %Columns
\usepackage[utf8]{inputenc}
\usepackage{caption}

\newcommand{\pp}[1]{{\color{black}{#1}}}

\title{Threshold Effects in Parameter Estimation from Compressed Data}
\author{Pooria Pakrooh,~\IEEEmembership{Student Member,~IEEE,}  Louis L. Scharf,~\IEEEmembership{Life Fellow,~IEEE}, \pp{and} Ali~Pezeshki,~\IEEEmembership{Member,~IEEE} \vfill\vfill
\thanks{This work is supported in part by NSF under grants CCF-1018472 and CCF-1422658.} \thanks{A preliminary version of a subset of the results reported here \pp{was} presented at the 2013 IEEE Global Conference on Signal and Information Processing (GlobalSIP), Austin, Tx, Dec. 3-5, 2013.}
\thanks{P. Pakrooh is with the Department of Electrical and
Computer Engineering, Colorado State University, Fort Collins, CO 80523, USA (e-mail: Pooria.Pakrooh@colostate.edu).}
\thanks{L. L. Scharf is with the Department of Mathematics and the Department of Statistics, Colorado State University, Fort Collins, CO 80523, USA (e-mail: Louis.Scharf@colostate.edu).}
\thanks{A. Pezeshki is with the Department of Electrical and
Computer Engineering and the Department of Mathematics, Colorado State University, Fort Collins, CO 80523, USA (e-mail: Ali.Pezeshki@colostate.edu).}

\thanks{}
\thanks{}
\thanks{}
\thanks{}}\newcommand\remark[1]{\textbf{Remark #1:}}
\DeclareMathOperator*{\argmin}{arg\,min}

\newcommand{\yb}{{\bf y}}

\newcommand{\thetab}{{\boldsymbol{\theta}}}

\newcommand{\xb}{{\bf x}}

\newcommand{\zb}{{\bf z}}
\newcommand{\Phib}{{\bf \Phi}}

\newcommand{\wb}{{\bf w}}
\newcommand{\Wb}{{\bf W}}

\begin{document}

\maketitle

\begin{abstract}

In this paper, we investigate threshold effects associated
with swapping of signal and noise subspaces in estimating
signal parameters from compressed noisy data. The term threshold
effect refers to a sharp departure of mean-squared error from the Cram\'{e}r-Rao bound
when the signal-to-noise ratio falls below a threshold SNR. In
many cases, the threshold effect is caused by a subspace swap
event, when the measured data (or its sample covariance) is
better approximated by a subset of components of an orthogonal
subspace than by the components of a signal subspace. We derive
analytical lower bounds on the probability of a subspace swap
in compressively measured noisy data. These bounds guide our
understanding of threshold effects and performance breakdown
for parameter estimation using compression. As a case study, we
investigate threshold effects in maximum likelihood (ML) estimation of directions
of arrival of two closely-spaced sources using co-prime subsampling. Our results show the
impact of compression on threshold SNR. A rule of thumb is that every doubling of compression ratio brings a penalty in threshold SNR of $3$ dB.
\end{abstract}

%\begin{keywords}
%Co-prime sampling, Cram\'{e}r-Rao bound, maximum likelihood estimation, mean squared
%error, random compression, subspace
%swap, threshold effects  \vspace{-.2cm}
%\end{keywords}

\section{Introduction}
The performance of many high resolution parameter estimation methods, including subspace and maximum likelihood methods, may suffer from performance breakdown, where the mean squared error (MSE) departs sharply from the Cram\'{e}r-Rao bound at low signal-to-noise ratio (SNR). Performance breakdown may happen when either the sample size or SNR falls below a certain threshold \cite{TKV}. The main reason for this threshold effect is that in low SNR or sample size regimes, parameter estimation methods lose their capability to resolve signal and noise subspaces. As a result of this, one or more components in the orthogonal (noise) subspace better approximate the data than at least one component of the signal subspace, which in turn leads to a large error in parameter estimation \cite{Scharf-SVD}. This phenomenon is called a subspace swap.

In this paper, we address the effect of compression on the probability of a subspace swap. In other words, we ask what effect compression has on the threshold SNR at which performance breaks down. To answer this question, we derive a lower bound on the probability of a subspace swap in parameter estimation from compressed noisy measurements. We consider two measurement models. In the first-order model, the parameters to be estimated modulate the mean of a complex multivariate normal set of measurements. In the second-order model, the parameters modulate the covariance of complex multivariate measurements. For these models, we derive analytical lower bounds on the probability of a subspace swap in compressively measured noisy data. These bounds guide our understanding of threshold effects and performance breakdown for parameter estimation using compression. As a case study, we investigate threshold effects in the maximum likelihood estimation of directions of arrival of two closely-spaced sources, using co-prime subsampling of a uniform line array. Co-prime sensor array processing was introduced recently by Vaidyanathan and Pal \cite{coprime1}-\nocite{coprime2}\cite{coprime3} as a sparse alternative to uniform line arrays. The concept was extended to sampling in multiple dimensions in subsequent papers by the same authors. In one dimension, the idea is to employ two uniform line arrays with spacings of $m_1$ and $m_2$ in units of half-wavelength, where $m_1$ and $m_2$ are co-prime. Our results show the impact of compression on threshold SNR, and can be used as a tool to predict the threshold SNR for different compression regimes in maximum likelihood estimation. Our simulation results indicate that compression brings a cost of about $10\mathrm{log}_{10}C~dB$ in threshold SNR, where $C$ is the compression ratio.

Co-prime sampling is just one type of sampling to which our results apply. Our probability bounds are tail probabilities of $F$-distributions, and they apply to any deterministic linear compression.

Other studies have also addressed the performance breakdown regions of high resolution parameter estimation methods. In \cite{TKV}, approximation of the probability of a subspace swap in the Singular Value Decomposition (SVD) is investigated. In \cite{Scharf-SVD} lower bounds on the probability of a subspace swap are derived for the problem of modal analysis. In \cite{Stoica2001} a lower bound on the probability of a subspace swap is derived by considering the separation of the estimates of signal and noise eigenvalues. In \cite{Shaghaghi15}, the authors propose a method to reduce the subspace leakage for the direction of arrival (DOA) estimation problem using root-MUSIC algorithm \cite{Barabell83}. In \cite{xu2002}-\nocite{Liu2008}\cite{Steinwandt14}, perturbation analysis of the SVD is carried out to study the performance of subspace based methods for parameter estimation, when \emph{subspace leakage} happens between signal and noise subspaces. Performance breakdown of maximum likelihood has been studied in \cite{Rife74}-\nocite{Steinhardt85}\nocite{Quinn94}\nocite{james95}\cite{Kaveh85} by perturbation analysis using an asymptotic assumption on the number of snapshots. More perturbation analysis may be found in the papers by Vaccaro \emph{et al.} \cite{Vaccaro94}-\nocite{Vaccaro87}\cite{Vaccaro93}. In \cite{Abra}, performance breakdown regions have been studied in the DOA estimation problem using asymptotic assumptions on the number of antennas and number of samples. It is shown that while a subspace swap is the main source of performance breakdown in maximum likelihood, earlier breakdown of MUSIC is due to the loss of resolution in separating closely-spaced sources.

%Subspace methods are widely used for high resolution parameter estimation. However, these methods suffer from performance breakdown, where the mean squared error (MSE) increases dramatically at low SNR. Performance breakdown may happen when either the sample size or signal-to-noise ratio (SNR) falls below a certain threshold \cite{TKV}. The main reason for this threshold effect is that in low SNR or sample size regimes, subspace methods lose their capability to resolve signal and noise subspaces. As a result of this, one or more components in the orthogonal (noise) subspace better approximate the data than at least one component of the signal subspace, which in turn leads to a large error in parameter estimation \cite{Scharf}. This phenomenon is called a subspace swap.\\\indent

%In \cite{TKV}, approximation of the probability of a subspace swap in the Singular Value Decomposition (SVD) is investigated. Ref. \cite{Scharf} extends the work of \cite{TKV} to lower bound the probability of a subspace swap. In \cite{Abra} the performance breakdown regions have been studied in the DOA estimation problem using asymptotic assumptions on the number of antennas and number of samples. It is shown that subspace swap and loss of resolution in separating closely spaced sources is responsible for performance breakdown of Maximum Likelihood (ML) and MUSIC Methods.\\\indent

%For our numerical results, we consider DOA estimation of two closely spaced sources and investigate the effect of compression with co-prime arrays \cite{coprime1, coprime2} on the probability of a subspace swap.

\section{Measurement Model}\label{sec:datamodel}
In the following subsections, we consider two models for the random measurement vector $\mathbf{y}\in \mathbb{C}^n$. In the first-order model, the parameters to be estimated nonlinearly modulate the mean of a complex multivariate normal vector, and in the second-order model the parameters nonlinearly modulate the covariance of a complex multivariate normal vector.
\subsection{Parameterized Mean Case}\label{sec:model_mean}
Let $\mathbf{y}\in \mathbb{C}^n$ be a complex  measurement vector in a signal plus noise model $\mathbf{y}=\mathbf{x}(\boldsymbol{\theta})+\mathbf{n}$. Here, we assume that $\mathbf{n}$ is a proper complex white Gaussian noise with covariance  $\sigma^2\mathbf{I}$ and $\mathbf{x}(\boldsymbol{\theta})$ is parameterized by $\boldsymbol{\theta}\in \mathbb{C}^p$, $p\leq n$. We assume that the parameters are nonlinearly embedded in $\xb(\thetab)$ as   $\mathbf{x}(\boldsymbol{\theta})=\mathbf{K}(\thetab)\boldsymbol{\alpha}$, where the columns of $\mathbf{K}(\thetab)=[{\mathbf{k}}(\boldsymbol{\theta}_1)\hspace{2mm} {\mathbf{k}}(\boldsymbol{\theta}_2) \hspace{2mm}\cdots\hspace{2mm} {\mathbf{k}}(\boldsymbol{\theta}_p)]$ define the signal subspace, and $\boldsymbol{\alpha}\in\mathbb{C}^p$ is a deterministic vector associated with the mode weights. Therefore, $\mathbf{y}$ is distributed as $\mathcal{CN}_n(\mathbf{K}(\thetab)\boldsymbol{\alpha},\sigma^2\mathbf{I})$, and the parameters $\boldsymbol{\theta}\in \mathbb{C}^p$ to be estimated nonlinearly modulate the mean of a complex multivariate normal vector. Assume we compress the measurement vector $\mathbf{y}$ by a unitary compression matrix $\mathbf{\Psi}=(\Phib\Phib^H)^{-1/2}\Phib$, where $\Phib\in\mathbb{C}^{m\times n},\  p\leq m<n$. Then, we obtain $\mathbf{w}=\mathbf{\Psi}\yb$ which is distributed as $\mathcal{CN}_m(\mathbf{z}(\boldsymbol{\theta}),\sigma^2\mathbf{I})$, where ${\mathbf{z}}(\boldsymbol{\theta})=\mathbf{\Psi}\mathbf{x}(\boldsymbol{\theta})$. We form the data matrix $\Wb=[\wb_1 \hspace{2mm} \wb_2\hspace{2mm} \cdots\hspace{2mm} \wb_M]$, where $\wb_i$'s are independent realizations of $\wb$. To specify a basis for the signal subspace and the orthogonal subspace in our problem, we define ${\mathbf{H}}(\thetab)=\mathbf{\Psi} \mathbf{K}(\thetab)=[{\mathbf{h}}(\boldsymbol{\theta}_1)\hspace{2mm} {\mathbf{h}}(\boldsymbol{\theta}_2) \hspace{2mm}\cdots\hspace{2mm} {\mathbf{h}}(\boldsymbol{\theta}_p)]$, with ${\mathbf{h}}(\boldsymbol{\theta}_i)=\mathbf{\Psi}\mathbf{k}(\boldsymbol{\theta}_i)$. The singular value decomposition of ${\mathbf{H}}_{m\times p} ~(p\leq m)$ is
\begin{equation}
{\mathbf{H}}=\mathbf{U}{\mathbf{\Sigma}}\mathbf{V}^H
\end{equation}
where
\begin{align}\label{eq:SVD}
\mathbf{U}&\in\mathbb{C}^{m\times m}: \mathbf{U}\mathbf{U}^H=\mathbf{U}^H\mathbf{U}=\mathbf{I}\nonumber\\
\mathbf{V}&\in\mathbb{C}^{p\times p}: \mathbf{V}\mathbf{V}^H=\mathbf{V}^H\mathbf{V}=\mathbf{I}\nonumber\\
\boldsymbol{\Sigma}&\in\mathbb{C}^{m\times p}:\boldsymbol{\Sigma}=\begin{bmatrix}
\boldsymbol{\Sigma}_p\\
\mathbf{0}
\end{bmatrix}\nonumber\\
\boldsymbol{\Sigma}_p&=\makebox{diag}(\sigma_1,\sigma_2,...,\sigma_p),~\sigma_1\geq\sigma_2\geq...\geq\sigma_p.
\end{align}
Now we can define the basis vectors from $\mathbf{U}=[\mathbf{u}_1,\mathbf{u}_2,...,\mathbf{u}_p|\mathbf{u}_{p+1},...,\mathbf{u}_m]=[\mathbf{U}_p|\mathbf{U}_0]$,
where $\langle\mathbf{U}_p\rangle$ and $\langle\mathbf{U}_0\rangle$ represent signal and orthogonal subspaces, respectively. The columns of $\mathbf{U}_p$ and  $\mathbf{U}_0$ can be considered as basis vectors for the signal and orthogonal subspaces, respectively.

\subsection{Parameterized Covariance Case}\label{sec:model_cov}

Assume in the signal plus noise model $\mathbf{y}=\mathbf{x}+\mathbf{n}$, the signal component $\mathbf{x}$ is of the form $\mathbf{x}=\mathbf{K}(\thetab)\boldsymbol{\alpha}$, where the columns of ${\mathbf{K}}(\thetab)=[{\mathbf{k}}(\boldsymbol{\theta}_1)\hspace{2mm} {\mathbf{k}}(\boldsymbol{\theta}_2) \hspace{2mm}\cdots\hspace{2mm} {\mathbf{k}}(\boldsymbol{\theta}_p)]$ are the modes and $\boldsymbol{\alpha}\in\mathbb{C}^p$ is the vector associated with the random mode weights. We assume $\boldsymbol{\alpha}$ is distributed as $\mathcal{CN}_p(0,\mathbf{R}_{\boldsymbol{\alpha}\boldsymbol{\alpha}})$. Therefore, $\mathbf{R}_{xx}(\thetab)=\mathbf{K}(\thetab)\mathbf{R}_{\boldsymbol{\alpha\alpha}}\mathbf{K}^H(\thetab)$ is parameterized by $\boldsymbol{\theta}\in \mathbb{C}^p$. We assume $\mathbf{n}$ is a proper complex white Gaussian noise with covariance  $\sigma^2\mathbf{I}$, and $\mathbf{x}$ and $\mathbf{n}$ are independent. Therefore, $\mathbf{y}$ is distributed as $\mathcal{CN}_n(\mathbf{0},\mathbf{R}_{\mathbf{yy}}(\boldsymbol{\theta}))$, where $\mathbf{R}_{\mathbf{yy}}(\boldsymbol{\theta})=\mathbf{K}(\thetab)\mathbf{R}_{\boldsymbol{\alpha\alpha}}\mathbf{K}^H(\thetab)+\sigma^2\mathbf{I}$. Such a data model arises in many applications such as direction of arrival and spectrum estimation.
%Assuming $\mathbf{R}_{xx}$ is rank deficient, we can write its eigenvalue decomposition as:
%\begin{equation}
%\mathbf{R}_{xx}=\mathbf{U}\boldsymbol{\Lambda}_{xx}\mathbf{U}^H
%\end{equation}
%where $\mathbf{U}$ and $\boldsymbol{\Lambda}_{xx}$ are defined as below:
%\begin{align}\label{eq:2}
%\mathbf{U}&\in\mathbb{C}^{n\times n}: \mathbf{U}\mathbf{U}^H=\mathbf{U}^H\mathbf{U}=\mathbf{I}\nonumber\\
%\boldsymbol{\Lambda}_{xx}&\in\mathbb{C}^{n\times n}:\boldsymbol{\Lambda}_{xx}=\begin{bmatrix}
%\boldsymbol{\Lambda}_p& \mathbf{0}\\
%\mathbf{0}& \mathbf{0}
%\end{bmatrix}\nonumber\\
%\boldsymbol{\Lambda}_p&=diag(\lambda_1,\lambda_2,...,\lambda_p),~\lambda_1\geq\lambda_2\geq...\geq\lambda_p>0
%\end{align}
%The unitary matrix $\mathbf{U}$ can be written as $\mathbf{U}=[\mathbf{u}_1,\mathbf{u}_2,...,\mathbf{u}_p|\mathbf{u}_{p+1},...,\mathbf{u}_n]=[\mathbf{U}_p|\mathbf{U}_0]$. Here $\langle\mathbf{U}_p\rangle$ represents the signal subspace and $\langle\mathbf{U}_0\rangle$ represents the orthogonal subspace which completes $\mathbb{C}^{n\times n}$. Fig. 1 gives a geometrical representation of (\ref{eq:2}).

\begin{figure}[h]
\begin{center}
\noindent
  \includegraphics[width=2.1in]{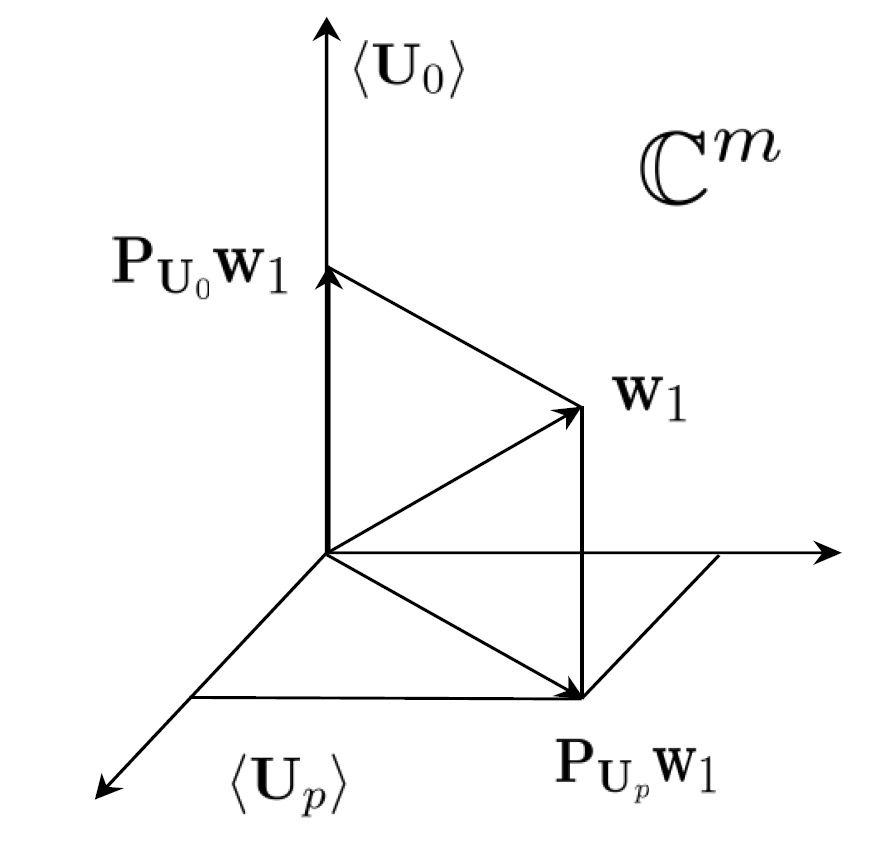}
  \caption{Signal and noise subspaces.}\label{Fig1Label}
\end{center}
\end{figure}

Assume we compress the measurement vector $\mathbf{y}$ by a unitary compression matrix $\mathbf{\Psi}=(\Phib\Phib^H)^{-1/2}\Phib$, where $\Phib\in\mathbb{C}^{m\times n} (m<n)$. Then, we obtain $\mathbf{w}=\mathbf{\Psi}\yb$ which is distributed as
\begin{equation}
\mathbf{w}\sim\mathcal{CN}_m(\mathbf{0},\mathbf{R}_{\mathbf{ww}})
\end{equation}
%Before proceeding, we first whiten $\tilde{\mathbf{y}}$ with $\mathbf{C}=(\mathbf{\Phi}\mathbf{\Phi}^H)^{-1/2}$ to produce $\hat{\mathbf{y}}=\mathbf{C}\tilde{\mathbf{y}}$. Therefore, $\hat{\mathbf{y}}$ is distributed as $\mathcal{CN}(\mathbf{0},\hat{\mathbf{R}})$ where
%\begin{equation}\label{eq:Rhat}
%\hat{\mathbf{R}}=\mathbf{C}\mathbf{\Phi}\mathbf{R}_{xx}(\boldsymbol{\theta})\mathbf{\Phi}^H\mathbf{C}^H+\sigma^2\mathbf{I}
%\end{equation}
where $\mathbf{R}_{\mathbf{ww}}=\mathbf{\Psi}\mathbf{K}(\thetab)\mathbf{R}_{\boldsymbol{\alpha\alpha}}\mathbf{K}^H(\thetab)\mathbf{\Psi}^H+\sigma^2\mathbf{I}$. We form the data matrix $\Wb=[\wb_1 \hspace{2mm} \wb_2\hspace{2mm} \cdots\hspace{2mm} \wb_M]$, where $\wb_i$'s are independent realizations of $\wb$. Each of these i.i.d. realizations consists of an i.i.d. realization of $\yb_i$, compressed by a common compressor $\mathbf{\Psi}$ for all $i=1,2,\dots,M$. We may define the signal covariance matrix after compression as
\begin{align}\label{eq:H_cov}
\mathbf{R}_{\mathbf{zz}}&=\mathbf{\Psi}\mathbf{K}(\thetab)\mathbf{R}_{\boldsymbol{\alpha\alpha}}\mathbf{K}^H(\thetab)\mathbf{\Psi}^H\nonumber\\
&=\mathbf{H}(\thetab)\mathbf{R}_{\boldsymbol{\alpha\alpha}}\mathbf{H}^H(\thetab),
\end{align}
where $\mathbf{H}(\thetab)=[{\mathbf{h}}(\boldsymbol{\theta}_1)\hspace{2mm} {\mathbf{h}}(\boldsymbol{\theta}_2) \hspace{2mm}\cdots\hspace{2mm} {\mathbf{h}}(\boldsymbol{\theta}_p)]$, and ${\mathbf{h}}(\boldsymbol{\theta}_i)=\mathbf{\Psi}{\mathbf{k}}(\boldsymbol{\theta}_i)$. Now, we can write the singular value decomposition of $\mathbf{R}_{\mathbf{zz}}$ and $\mathbf{R}_{\mathbf{ww}}$ as
\begin{align}\label{eq:5}
\mathbf{R}_{\mathbf{zz}}&=\mathbf{U}\boldsymbol{\Lambda}\mathbf{U}^H\nonumber\\
\mathbf{R}_{\mathbf{ww}}&=\mathbf{U}(\boldsymbol{\Lambda}+\sigma^2\mathbf{I})\mathbf{U}^H
\end{align}
where $\mathbf{U}$ and $\boldsymbol{\Lambda}$ are defined as
\begin{align}\label{eq:6}
\mathbf{U}&\in\mathbb{C}^{m\times m}: \mathbf{U}\mathbf{U}^H=\mathbf{U}^H\mathbf{U}=\mathbf{I}\nonumber\\
\boldsymbol{\Lambda}&\in\mathbb{C}^{m\times m}:\boldsymbol{\Lambda}=\begin{bmatrix}
\boldsymbol{\Lambda}_p& \mathbf{0}\\
\mathbf{0}& \mathbf{0}
\end{bmatrix}\nonumber\\
\boldsymbol{\Lambda}_p&=\makebox{diag}(\lambda_1,\lambda_2,...,\lambda_p),~\lambda_1\geq\lambda_2\geq...\geq\lambda_p.
\end{align}
Assuming $\mathbf{R}_{\mathbf{zz}}$ has rank $p$, the unitary matrix $\mathbf{U}$ can be written as $\mathbf{U}=[\mathbf{u}_1,\mathbf{u}_2,...,\mathbf{u}_p|\mathbf{u}_{p+1},...,\mathbf{u}_m]=[\mathbf{U}_p|\mathbf{U}_0]$. Here $\langle\mathbf{U}_p\rangle$ represents the signal subspace and $\langle\mathbf{U}_0\rangle$ represents the orthogonal subspace which completes $\mathbb{C}^{m\times m}$, assuming $p\leq m<n$. Figure \ref{Fig1Label} gives a geometrical representation of (\ref{eq:6}).

\section{Bound on the Probability of a Subspace Swap after Compression}\label{sec:bounds}
To bound the probability of a subspace swap for the compressed measurements $\Wb$, we define the following events:

\begin{itemize}
\item $E$ is the event that one or more modes of the orthogonal subspace resolve more energy in $\Wb$ than one or more modes of the noise-free signal subspace. Therefore, $E$ may be written as

\begin{equation}\label{eq:SS1}
E=\cup_{q=1}^{p}E(q),
\end{equation}
where $E(q)$ is the following subset of the subspace swap event $E$,
\begin{equation}\label{eq:SS_main}
\min_{\mathbf{A}\in\mathcal{I}_{p,q}}\mathrm{tr}(\Wb^H\mathbf{P}_{\mathbf{H}\mathbf{A}}\Wb)< \max_{\mathbf{B}\in \mathbb{C}^{(n-p)\times q}}\mathrm{tr}(\Wb^H\mathbf{P}_{\mathbf{U}_{0}\mathbf{B}}\Wb),
\end{equation}
and $\mathcal{I}_{p,q}$ is the set of all $p\times q$ slices of the identity matrix $\mathbf{I}_p$. Here, the columns of $\mathbf{H}$ are the modes defined in Section \ref{sec:datamodel}, and $\mathbf{A}$ selects $q$ of the columns of $\mathbf{H}$.

\item $F$ is the event that the average energy resolved in the orthogonal subspace $\langle {\mathbf{U}}_0 \rangle$ is greater than the average energy resolved in the noise-free signal subspace $\langle {\mathbf{U}}_p \rangle$ (or equivalently $\langle \mathbf{H} \rangle$). Then, the following bounds establish that $F$ is a subset of $E(1)$, which is in turn a subset of the swap event $E$:

\begin{align}
\min_{1\leq i\leq p}\mathrm{tr}(\Wb^H\mathbf{P}_{\mathbf{h}_i}\Wb)&\le \frac{1}{p}\mathrm{tr}({\Wb}^H{\mathbf{P}}_{{\mathbf{U}}_p}{\Wb})\nonumber\\
&<\frac{1}{m-p}\mathrm{tr}({\Wb}^H{\mathbf{P}}_{{\mathbf{U}}_0}{\mathbf{w}})\nonumber\\
&\le \max_{p+1\leq i\leq m} \mathrm{tr}(\Wb^H\mathbf{P}_{\mathbf{u}_i}\Wb)\nonumber\\
&\le \max_{\mathbf{b}\in \mathbb{C}^{(n-p)\times 1}}\mathrm{tr}(\Wb^H\mathbf{P}_{\mathbf{U}_{0}\mathbf{b}}\Wb).
\end{align}

\item $G$ is the event that the energy resolved in the apriori minimum mode $\mathbf{h}_{min}$ of the noise-free signal subspace $\langle \mathbf{H} \rangle$ (or equivalently $\langle {\mathbf{U}}_p \rangle$) is smaller than the average energy resolved in the orthogonal subspace $\langle {\mathbf{U}}_0 \rangle$. For the parameterized mean measurement model, we define $\mathbf{h}_{min}$ as
    \begin{equation}
    \mathbf{h}_{min}=\argmin_{\mathbf{h}\in\{\mathbf{h}(\thetab_1),\mathbf{h}(\thetab_2),\dots,\mathbf{h}(\thetab_p)\}}|\mathbf{h}^H\zb(\thetab)|^2,
    \end{equation}
and for the parameterized covariance measurement model as

 \begin{equation}\label{eq:hmin}
 \mathbf{h}_{min}=\argmin_{\mathbf{h}\in\{\mathbf{h}(\thetab_1),\mathbf{h}(\thetab_2),\dots,\mathbf{h}(\thetab_p)\}}|\mathbf{h}^H\mathbf{R}_{\zb\zb}(\thetab)\mathbf{h}|^2.
    \end{equation}
Then, the following bounds establish that $G$ is a subset of $E(1)$, which is in turn a subset of the swap event $E$:

\begin{align}
\min_{1\leq i\leq p}\mathrm{tr}(\Wb^H\mathbf{P}_{\mathbf{h}_i}\Wb)&\leq \mathrm{tr}(\Wb^H\mathbf{P}_{\mathbf{h}_{min}}\Wb)\nonumber\\
&<\frac{1}{m-p}\mathrm{tr}({\Wb}^H{\mathbf{P}}_{{\mathbf{U}}_0}{\mathbf{w}})\nonumber\\
&\le \max_{p+1\leq i\leq m} \mathrm{tr}(\Wb^H\mathbf{P}_{\mathbf{u}_i}\Wb)\nonumber\\
&\le \max_{\mathbf{b}\in \mathbb{C}^{(n-p)\times 1}}\mathrm{tr}(\Wb^H\mathbf{P}_{\mathbf{U}_{0}\mathbf{b}}\Wb).
\end{align}

\end{itemize}
Since events $F$ and $G$ are subsets of event $E$, their probabilities of occurrence give lower bounds on the probability of a subspace swap, $P_{ss}\triangleq P(E)$. We use these events to derive lower bounds on the probability of a subspace swap for the two data models given in Section \ref{sec:datamodel}.

%For the cases where signal subspace eigenvalues are clustered, event $F$ seems a better subset of a subspace swap event. On the other hand, when  signal subspace eigenvalues are dispersed, event $G$ seems a better subset of a subspace swap event \cite{Scharf-SVD}.
\subsection{Parameterized Mean Case}\label{subsec:mean}

For the parameterized mean measurement model discussed in Section \ref{sec:model_mean}, we start with event $F$ and define
%\footnote{By defining $\mathbf{T}_G=(m-p)^{-1}\mathbf{P}_{\mathbf{U}_0}-\mathbf{P}_{\mathbf{u}_{p}}
%$, the same line of argument can be made for event $G$.}
\begin{equation}\label{eq:7Mean}
\mathbf{T}_F=\frac{1}{m-p}\mathbf{P}_{\mathbf{U}_0}-\frac{1}{p}\mathbf{P}_{\mathbf{U}_{p}}
\end{equation}
where $\mathbf{P}_{\mathbf{U}_{p}}=\mathbf{U}_{p}\mathbf{U}_{p}^H$ is the orthogonal projection onto the signal subspace and $\mathbf{P}_{\mathbf{U}_0}=\mathbf{U}_0\mathbf{U}_0^H$ is the orthogonal projection onto the orthogonal (noise) subspace. According to the definition of event $F$ we can lower bound the probability of a subspace swap $P_{ss}$ as
\begin{align}\label{eq:8Mean}
P_{ss}\geq P(\mathrm{tr}\big[\Wb^H\mathbf{T}_F\Wb\big]>0)
\end{align}
Therefore, we have
\begin{align}\label{eq:9Mean}
P_{ss}&\geq P(\mathrm{tr}\big[\Wb^H\mathbf{T}_F\Wb\big]>0)\nonumber\\
&=P(\frac{\mathrm{tr}\big[\Wb^H\mathbf{U}_p\mathbf{U}_p^H\Wb\big]/2p}{\mathrm{tr}\big[\Wb^H\mathbf{U}_0\mathbf{U}_0^H\Wb\big]/2(m-p)}<1)\nonumber\\
&=P(\frac{\sum_{i=1}^M\|\mathbf{U}_p^H\wb_i\|_2^2/2p}{\sum_{i=1}^M\|\mathbf{U}_0^H\wb_i\|_2^2/2(m-p)}<1).
\end{align}
Here, the $\mathbf{U}_p^H\wb_i$ are independent and identically distributed as
\begin{equation}
\mathbf{U}_p^H\wb_i\sim\mathcal{CN}_p(\mathbf{U}_p^H\mathbf{z}(\boldsymbol{\theta}),\sigma^2\mathbf{I})~~\forall 1\leq i\leq M.
\end{equation}
Therefore, $\|\mathbf{U}_p^H\wb_i\|_2^2/\sigma^2\sim\mathcal{\chi}^2_{2p}(\|\mathbf{z}(\boldsymbol{\theta})\|_2^2/\sigma^2)$, which is the distribution of a complex noncentral chi-squared random variable with $2p$ degrees of freedom and noncentrality parameter $\|\mathbf{z}(\boldsymbol{\theta})\|_2^2/\sigma^2$. Also, since $\langle\mathbf{U}_p\rangle$ and $\langle\mathbf{U}_0\rangle$ are orthogonal, we can conclude that in (\ref{eq:9Mean}), each $\|\mathbf{U}_0^H\wb_i\|_2^2/\sigma^2$ is independent of $\|\mathbf{U}_p^H\wb_i\|_2^2/\sigma^2$ and is distributed as $\mathcal{\chi}^2_{2(m-p)}$. Hence, the term $\frac{\|\mathbf{U}_p^H\wb\|_2^2/2p}{\|\mathbf{U}_0^H\wb\|_2^2/2(m-p)}$ is the ratio of two independent normalized chi-squared random variables and is distributed as $F_{2pM,2(m-p)M}(\|\mathbf{z}(\boldsymbol{\theta})\|_2^2/\sigma^2)$, which is a  noncentral $F$ distribution with $2pM$ and $2(m-p)M$ degrees of freedom and noncentrality parameter $\|\mathbf{z}(\boldsymbol{\theta})\|_2^2/\sigma^2$. Thus, the probability of a subspace swap after compression is lower bounded by the probability that a $F_{2pM,2(m-p)M}(\|\mathbf{z}(\boldsymbol{\theta})\|_2^2/\sigma^2)$ distributed random variable is less than $1$. When there is no compression, this lower bound turns into the probability that a $F_{2pM,2(n-p)M}(\|\mathbf{x}(\boldsymbol{\theta})\|_2^2/\sigma^2)$ random variable is less than $1$.

For event $G$, we define
\begin{equation}
\mathbf{T}_G=\frac{1}{m-p}\mathbf{P}_{\mathbf{U}_0}-\mathbf{P}_{\mathbf{h}_{min}}.
\end{equation}
Here, we define $\boldsymbol{\rho}_{min}=\frac{\mathbf{h}_{min}}{\|\mathbf{h}_{min}\|_2}$. Therefore $\mathbf{P}_{\mathbf{h}_{min}}=\boldsymbol{\rho}_{min}\boldsymbol{\rho}_{min}^H$, and we have
\begin{align}\label{eq:10Mean}
P_{ss}&\geq P(\mathrm{tr}\big[\Wb^H\mathbf{T}_G\Wb\big]>0)\nonumber\\
&=P(\frac{\mathrm{tr}\big[\Wb^H\boldsymbol{\rho}_{min}\boldsymbol{\rho}_{min}^H\Wb\big]/2}{\mathrm{tr}\big[\Wb^H\mathbf{U}_0\mathbf{U}_0^H\Wb\big]/2(m-p)}<1)\nonumber\\
&=P(\frac{\sum_{i=1}^M\|\boldsymbol{\rho}_{min}^H\wb_i\|_2^2/2}{\sum_{i=1}^M\|\mathbf{U}_0^H\wb_i\|_2^2/2(m-p)}<1).
\end{align}
Here, we have
\begin{equation}
\mathbf{\boldsymbol{\rho}}_{min}^H\wb_i\sim\mathcal{CN}(\boldsymbol{\rho}_{min}^H\mathbf{z}(\boldsymbol{\theta}),\sigma^2\mathbf{I})~~\forall 1\leq i\leq M.
\end{equation}
Therefore, $\|\boldsymbol{\rho}_{min}^H\wb_i\|_2^2/\sigma^2\sim\mathcal{\chi}^2_{2}(|\boldsymbol{\rho}_{min}^H\mathbf{z}(\boldsymbol{\theta})|^2/\sigma^2)$ which is the distribution of a complex noncentral chi-squared random variable with $2$ degrees of freedom and noncentrality parameter $|\boldsymbol{\rho}_{min}^H\mathbf{z}(\boldsymbol{\theta})|^2/\sigma^2$. Thus, with the same type of arguments as for event $F$, we can conclude that the term $\frac{\sum_{i=1}^M\|\boldsymbol{\rho}_{min}^H\wb_i\|_2^2/2}{\sum_{i=1}^M\|\mathbf{U}_0^H\wb_i\|_2^2/2(m-p)}$ is distributed as $F_{2M,2(m-p)M}(|\boldsymbol{\rho}_{min}^H\mathbf{z}(\boldsymbol{\theta})|^2/\sigma^2)$, which is a  noncentral $F$ distribution with $2M$ and $2(m-p)M$ degrees of freedom and noncentrality parameter $|\boldsymbol{\rho}_{min}^H\mathbf{z}(\boldsymbol{\theta})|^2/\sigma^2$. When there is no compression, this turns into the probability that a $F_{2M,2(n-p)M}(|\boldsymbol{\kappa}_{min}^H\mathbf{x}(\boldsymbol{\theta})|^2/\sigma^2)$ random variable is less than $1$. Here, $\boldsymbol{\kappa}_{min}=\frac{\mathbf{k}_{min}}{\|\mathbf{k}_{min}\|_2}$, and $\mathbf{k}_{min}$ is the apriori minimum mode of the signal subspace before compression.
\subsection{Parameterized Covariance Case}\label{subsec:cov}
For the parameterized covariance measurement model discussed in Section \ref{sec:model_cov}, we start with event $F$. In this case, the columns of the measurement matrix $\mathbf{W}$ are i.i.d. random vectors distributed as $\mathcal{CN}(\mathbf{0},\mathbf{R}_{\wb\wb})$, and similar to the mean case we have
\begin{align}\label{eq:9Cov}
P_{ss}&\geq P(\mathrm{tr}\big[\Wb^H\mathbf{T}_F\Wb\big]>0)\nonumber\\
&=P(\frac{\mathrm{tr}\big[\Wb^H\mathbf{U}_p\mathbf{U}_p^H\Wb\big]/2p}{\mathrm{tr}\big[\Wb^H\mathbf{U}_0\mathbf{U}_0^H\Wb\big]/2(m-p)}<1)\nonumber\\
&=P(\frac{\sum_{i=1}^M\|\mathbf{U}_p^H\wb_i\|_2^2/2p}{\sum_{i=1}^M\|\mathbf{U}_0^H\wb_i\|_2^2/2(m-p)}<1).
\end{align}
Here, the $\mathbf{U}_p^H\wb_i$ are i.i.d. and distributed as
\begin{equation}
\mathbf{U}_p^H\wb_i\sim\mathcal{CN}_p(\mathbf{0},\mathbf{\Lambda}_p+\sigma^2\mathbf{I}_p)~~\forall 1\leq i\leq M.
\end{equation}
Therefore we can write
\begin{equation}
\|\mathbf{U}_p^H\wb_i\|_2^2=\sum_{i=1}^{p}(\lambda_i+\sigma^2)\rho_i,
\end{equation}
where $\rho_i$'s are i.i.d. random variables, each distributed as ${\chi}^2_2$. Therefore,

\begin{equation}
\sum_{i=1}^M\|\mathbf{U}_p^H\wb_i\|_2^2=\sum_{i=1}^{p}(\lambda_i+\sigma^2)\xi_i,
\end{equation}
where $\xi_i$'s are i.i.d. random variables, each distributed as ${\chi}^2_{2M}$. Also, we can write $\sum_{i=1}^M\|\mathbf{U}_0^H\wb_i\|_2^2=\sigma^2\nu$, where $\nu$ is distributed as ${\chi}^2_{2M(m-p)}$ and is independent of the $\xi_i$'s. Therefore, we have

\begin{align}\label{eq:10Cov}
P_{ss}&\geq P(\frac{\sum_{i=1}^M\|\mathbf{U}_p^H\wb_i\|_2^2/2p}{\sum_{i=1}^M\|\mathbf{U}_0^H\wb_i\|_2^2/2(m-p)}<1)\nonumber\\
&=P(\frac{\sum_{i=1}^{p}(1+\lambda_i/\sigma^2)\xi_i/2Mp}{\nu/2M(m-p)}<1).
\end{align}
Here, the term $\frac{\sum_{i=1}^{p}(1+\lambda_i/\sigma^2)\xi_i/2Mp}{\nu/2M(m-p)}$ is distributed as $GF\big[(1+\frac{\lambda_1}{\sigma^2}),\dots,(1+\frac{\lambda_p}{\sigma^2});2Mp;2M(m-p)\big]$, which is the distribution of a generalized $F$ random variable \cite{GFdunk}. Thus, the probability of a subspace swap in this case is lower bounded by the probability that a $GF\big[(1+\frac{\lambda_1}{\sigma^2}),\dots,(1+\frac{\lambda_p}{\sigma^2});2M;2M(m-p)\big]$ random variable is less than $1$. Without compression, this turns into the probability that a $GF\big[(1+\frac{\tilde{\lambda}_1}{\sigma^2}),\dots,(1+\frac{\tilde{\lambda_p}}{\sigma^2});2Mp;2M(m-p)\big]$ random variable is less than $1$. Here $\tilde{\lambda_i}$'s are the eigenvalues of the signal covariance matrix $\mathbf{R}_{\xb\xb}$ before compression.

We can also derive the probability of the event $G$ for the parameterized covariance measurement model. In this case we have
\begin{align}\label{eq:10Cov}
P_{ss}&\geq P(\mathrm{tr}\big[\Wb^H\mathbf{T}_G\Wb\big]>0)\nonumber\\
&=P(\frac{\sum_{i=1}^M\|\boldsymbol{\rho}_{min}^H\wb_i\|_2^2/2}{\sum_{i=1}^M\|\mathbf{U}_0^H\wb_i\|_2^2/2(m-p)}<1),
\end{align}
where $\boldsymbol{\rho}_{min}=\frac{\mathbf{h}_{min}}{\|\mathbf{h}_{min}\|_2}$, and $\mathbf{h}_{min}$ is the apriori minimum mode of the signal subspace given by (\ref{eq:hmin}). Here, $\boldsymbol{\rho}_{min}^H\wb_i$ is distributed as
\begin{equation}
\boldsymbol{\rho}_{min}^H\wb_i\sim\mathcal{CN}(0,\tau)~~\forall 1\leq i\leq M,
\end{equation}
where $\tau=\boldsymbol{\rho}_{min}^H\mathbf{R}_{\wb\wb}\boldsymbol{\rho}_{min}$. Therefore,
\begin{equation}
\sum_{i=1}^M\|\boldsymbol{\rho}_{min}^H\wb_i\|_2^2/\tau\sim\chi^2_{2M},
\end{equation}
and we have
\begin{align}\label{eq:11Cov}
P_{ss}&\geq P(\frac{\sum_{i=1}^M\|\boldsymbol{\rho}_{min}^H\wb_i\|_2^2/2}{\sum_{i=1}^M\|\mathbf{U}_0^H\wb_i\|_2^2/2(m-p)}<1)\nonumber\\
&=P(\vartheta<\frac{\sigma^2}{\tau}),
\end{align}
where $\vartheta$  is distributed as $F_{2M,2M(m-p)}$, which is a central $F$ random variable with $2M$ and $2M(m-p)$ degrees of freedom. Without compression, this turns into the probability that a $F_{2M,2M(n-p)}$ random variable is less than $\frac{\sigma^2}{\tilde{\tau}}$, where ${\tilde{\tau}}=\boldsymbol{\kappa}_{min}^H\mathbf{R}_{\yb\yb}\boldsymbol{\kappa}_{min}$, $\boldsymbol{\kappa}_{min}=\frac{\mathbf{k}_{min}}{\|\mathbf{k}_{min}\|_2}$, and $\mathbf{k}_{min}$ is the apriori minimum mode of the signal subspace before compression.

\remark{1} In Sections (\ref{subsec:mean}) and (\ref{subsec:cov}), we have derived lower bounds on the probability of a subspace swap for the case that $\mathbf{\Psi}=(\Phib\Phib^H)^{-1/2}\Phib$ is deterministic, as in standard or co-prime subsamplings. In  the case that $\mathbf{\Psi}$ is random, these probability bounds would have to be integrated over the distribution of $\mathbf{\Psi}$ to give lower bounds on marginal probabilities of a subspace swap. For example, for random $\mathbf{\Psi}$ and for the subevent $F$ we have
\begin{equation}
P_{ss}=\int{P(E|\mathbf{\Psi})P(\mathbf{\Psi})d\mathbf{\Psi}}\ge\int{P(F|\mathbf{\Psi})P(\mathbf{\Psi})d\mathbf{\Psi}}
\end{equation}
where $P(F|\mathbf{\Psi})$ is given in Sections (\ref{subsec:mean}) and (\ref{subsec:cov}) for the parameterized mean and parameterized covariance measurement models, respectively. For the class of random compression matrices that have density functions of the form $g(\Phib\Phib^H)$, that is, the distribution of $\Phib$ is right orthogonally invariant, $\mathbf{\Psi}$ is uniformly distributed on the Stiefel manifold $\mathcal{V}_{m}(\mathbb{C}^n)$ \cite{Chikuse2003}. The compression matrix $\Phib$ whose elements are i.i.d. standard normal random variables is one such matrix.
\section{Simulation Results}\label{sec:sr}

In this section, we present numerical examples to show the impact of compression on threshold effects for estimating directions of arrival using a sensor array. We consider a dense uniform line array with $n$ elements at half-wavelength inter-element spacings. We compress this array to $m$ dimensions using co-prime subsampling. In co-prime compression, we uniformly subsample the dense array once by a factor $m_1$ and once by a factor $m_2$, where $m_1$ and $m_2$ are co-prime. We then interleave these two subarrays to form the co-prime array of $m_1+2m_2-1$ elements. We note that although we are compressing the array by a factor $n/m$ for the co-prime array, the dense and the compressed arrays still have the same total aperture. The geometry of the dense and co-prime arrays are shown in Figure \ref{fig:copgeo}. We consider two point sources at far field at electrical angles $\theta_1=0$ and $\theta_2=\pi/n$. We set the amplitudes of these sources $\alpha_1=\alpha_2=1$. The Rayleigh limit of the dense array in electrical angle is $2\pi/n$. Therefore, in our examples the two sources are separated by half the Rayleigh limit of the dense array. We present the results for the parameterized mean and parameterized covariance cases.
 \begin{figure}[!ht]\centering
               \includegraphics[width=0.7\linewidth]{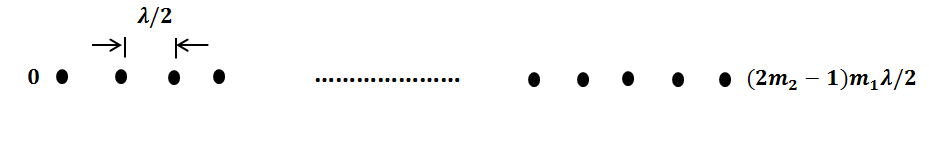}
                   \vspace{-0.6cm}\caption*{(a)}\vspace{2mm}
\end{figure}
 \begin{figure}[!ht]\centering
              \hspace{-1.2in} \includegraphics[width=0.52\linewidth]{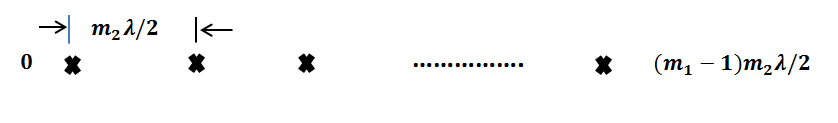}
                   \vspace{-0.6cm}\caption*{(b)}\vspace{2mm}
\end{figure}
 \begin{figure}[!ht]\centering
               \hspace{0.3in}\includegraphics[width=0.56\linewidth]{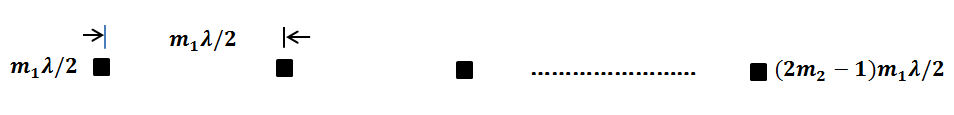}
                   \vspace{0cm}\caption*{(c)}

               \end{figure}
\begin{figure}[!ht]\centering
                   \caption{Geometry of the dense array (a), and co-prime subarrays (b), (c). At $m_1=11$ and $m_2=9$, $(2m_2-1)m_1\lambda/2=187\lambda/2$.}\label{fig:copgeo}
\end{figure}

\subsection{Parameterized Mean Case}

\begin{figure}[!ht]\centering
\includegraphics[width=320pt]{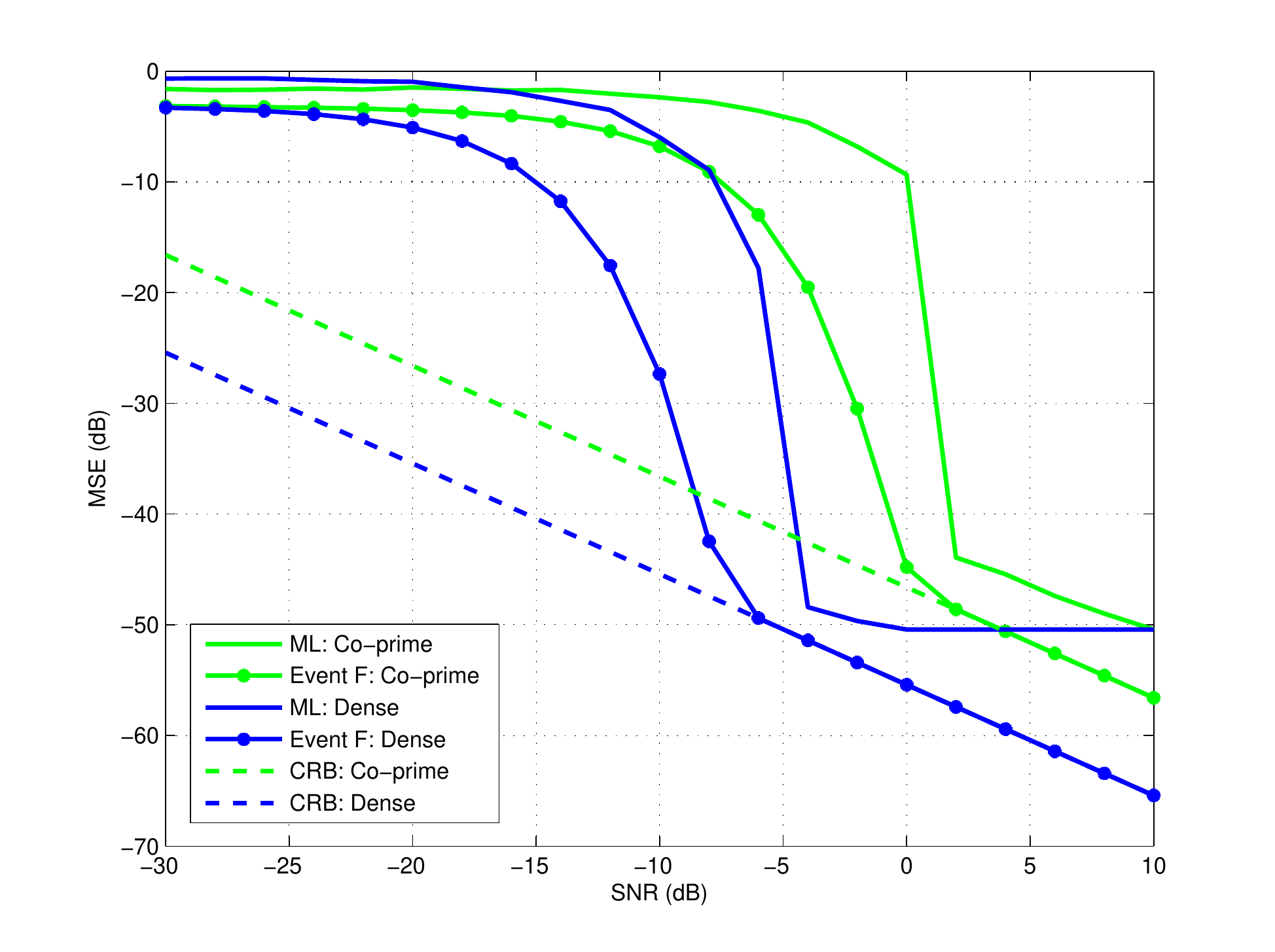}
  \caption{Parameterized mean case. Dense $188$ element array and $28$ element co-prime array. MSE bounds and MSE for ML estimation of $\theta_1=0$ in the presence of an interfering source at $\theta_2=\pi/188$; $200$ trials.}\label{fig:MSE_Mean}
\end{figure}

In this case, the Monte Carlo simulation consists of $200$ independent realizations of the measurement vector $\yb$ for a dense array of $188$ elements, each for a single snapshot ($M=1$). Then we compress these measurements to simulate the data for the co-prime compressed array of $28$ elements with $m_1=11$ and $m_2=9$. The compression ratio is $\frac{n}{m}\approx6.7$. Figure \ref{fig:MSE_Mean} shows the MSE for the maximum likelihood estimator of the source at $\theta_1$ in the presence of the interfering source at $\theta_2$. The CRB corresponding to the $188$-element dense array is also shown in this figure as a reference for performance analysis. Figure \ref{fig:MSE_Mean} also shows approximations to the MSE (in starred solid lines) obtained using the \emph {method of intervals} (introduced in \cite{VanTrees} and used in \cite{Scharf-SVD}). At each SNR, the approximate MSE $\sigma_T^2$ is computed as

\begin{equation}\label{eq:MI}
\sigma_T^2=P_{ss}\sigma_0^2+(1-P_{ss})\sigma_{CR}^2.
\end{equation}
Here, $P_{ss}$ is the probability of the subspace swap as a function of SNR, which we approximate using the lower bound in (\ref{eq:8Mean}); $\sigma_{CR}^2$ is the value of the CRB as a function of SNR, and $\sigma_0^2$ is the variance of the error given the occurrence of a subspace swap. The justification for using this formula is that when a subspace swap does not occur, MSE almost follows the CRB . However, given the occurrence of the subspace swap (and in the absence of any prior knowledge) the error in estimating the electrical angle $\theta_1$ may be taken to be uniformly distributed between $(-\pi/2,\pi/2)$ and the error variance is $\sigma_0^2=\pi^2/12$.

\begin{figure}[!ht]\centering
\includegraphics[width=320pt]{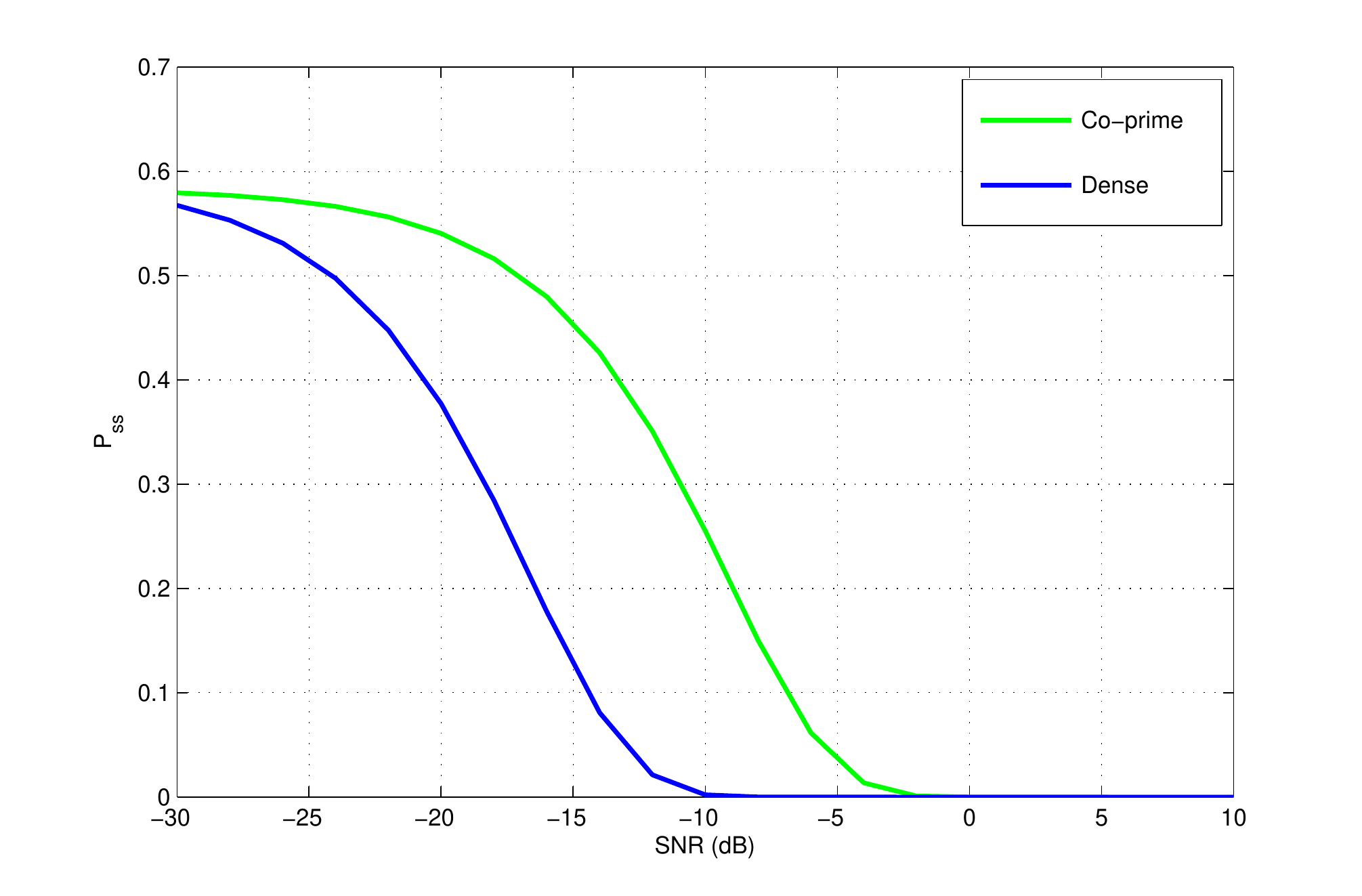}
  \caption{Parameterized mean case. Analytical lower bounds (event $F$) for the probability of subspace swap for estimation of the angle of a source at $\theta_1=0$ in the presence of an interfering source at $\theta_2=\pi/188$ using $188$ element dense array and $28$ element co-prime array.}\label{fig:Prob_Mean}
\end{figure}

Figure \ref{fig:MSE_Mean} shows that performance loss, measured by onset of threshold effect is approximately $10\mathrm{log}_{10}n/m$. Our approximations on MSE also predict the same SNR difference in the onset of the performance breakdown. Figure \ref{fig:Prob_Mean} shows our bounds on the probability of a subspace swap for the dense and co-prime arrays which are obtained using event $F$ in Section \ref{sec:bounds}. The ML curves of Figure \ref{fig:MSE_Mean} would approach the CRB at high SNR were it not for the quantization of our ML simulation code.

%The curves in Figure \ref{fig:Prob_Mean} are exact probabilities. The curves of Figure \ref{fig:MSE_Mean} are Monte-Carlo approximations for dense and co-prime ML, and interval approximations for the bounds, using event $F$.

\subsection{Parameterized Covariance Case}

\begin{figure}[!ht] \centering
\includegraphics[width=320pt]{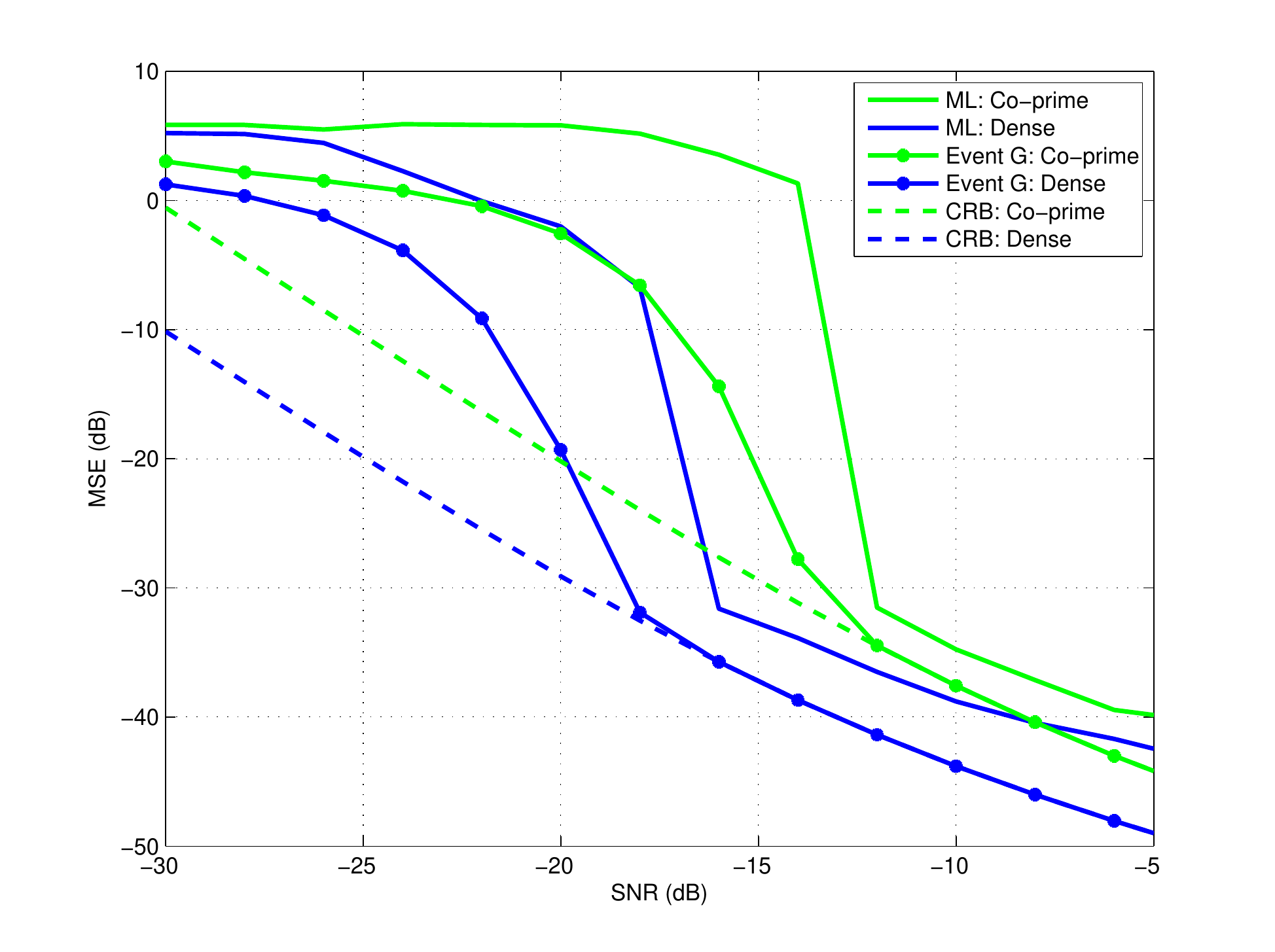}
  \caption{Parameterized covariance case. Dense $36$ element array and $12$ element co-prime array. MSE bounds and MSE for ML estimation of $\theta_1=0$ in the presence of an interfering source at $\theta_2=\pi/36$; $200$ snapshots and $200$ trials.}\label{fig:Covariance_MSE}
\end{figure}

\begin{figure}[!ht]\centering
\includegraphics[width=320pt]{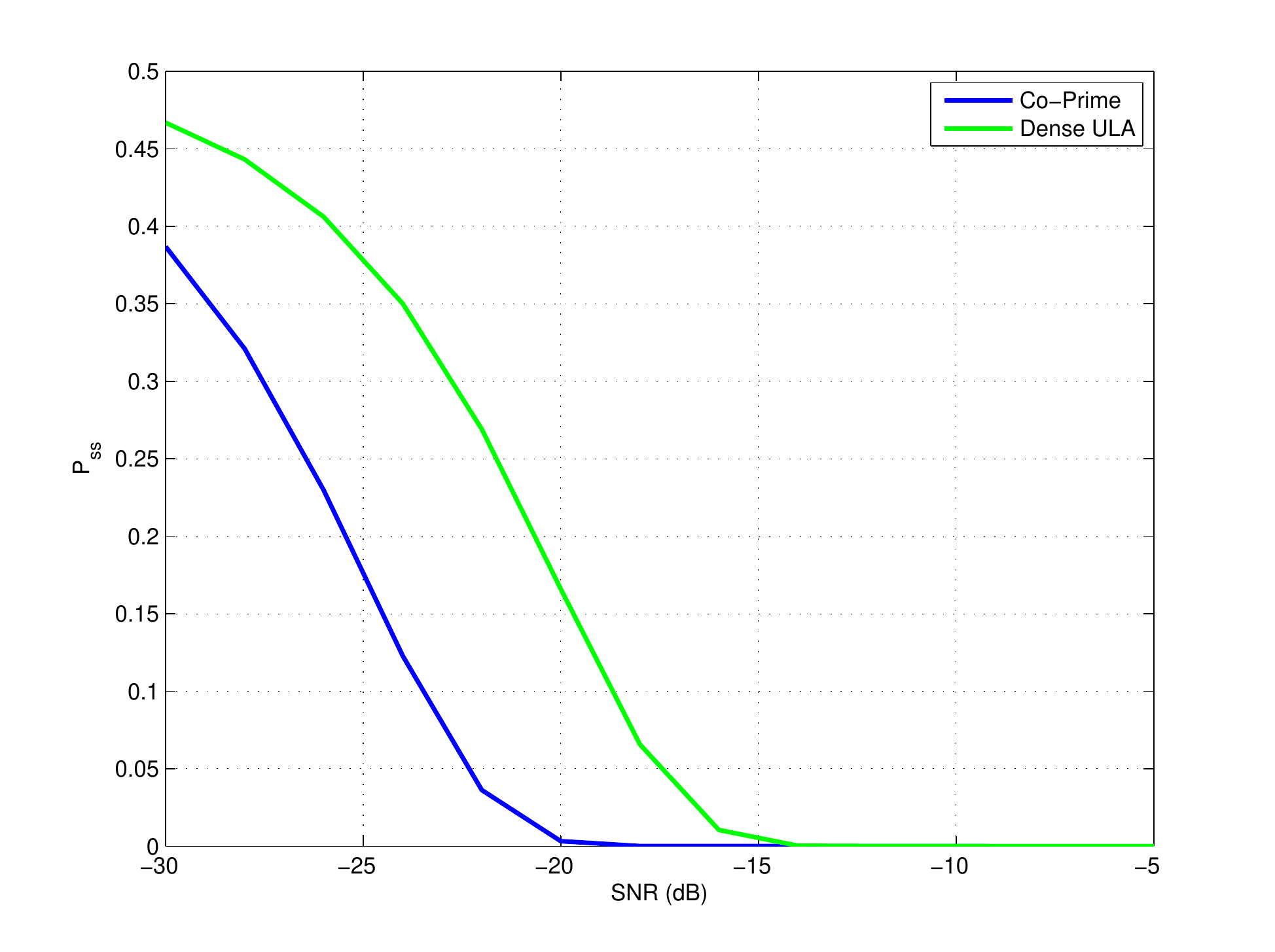}
  \caption{Parameterized covariance case. Analytical lower bounds (event $G$) for the probability of a subspace swap using co-prime compression for the estimation of the angle of a source at $\theta_1=0$ in the presence of an interfering source at $\theta_2=\pi/36$ using $36$ element dense array and $12$ element co-prime array.}\label{fig:Prob_Cov}
\end{figure}

We conduct the same set of Monte Carlo simulations for the stochastic data model. Here we draw $M=200$ independent snapshots for a dense array of $36$ elements over $200$ independent realizations, and compress them to simulate the data for the co-prime array of $12$ elements with $m_1=5$ and $m_2=4$. The compression ratio is $\frac{n}{m}=3$. Figure \ref{fig:Covariance_MSE} shows the results for the MSE of the maximum likelihood estimator of the source at $\theta_1=0$ in the presence of the interfering source at $\theta_2=\pi/36$. Our approximations for the MSE using the \emph {method of intervals} in (\ref{eq:MI}) and the Cram\'{e}r-Rao bound are also shown for each array. Figure \ref{fig:Covariance_MSE} shows that performance loss, measured by onset of threshold effect is approximately $10\mathrm{log}_{10}n/m$. Our bounds on the probability of a subspace swap using event $G$ in Section \ref{sec:bounds} are shown in Figure \ref{fig:Prob_Cov} for the dense and co-prime arrays.

\section{Conclusion}

We have addressed the effect of compression on the probability of a subspace swap. A subspace swap is known to be the main source of performance breakdown in maximum likelihood parameter estimation, wherein one or more modes of a noise subspace better approximate a measurement than one or more modes of a signal subspace. We have derived an analytical bound on this probability for two measurement models. In the first-order model, the parameters modulate the mean of a set of complex i.i.d. multivariate normal measurements. In the second-order model, the parameters to be estimated modulate a covariance matrix. Our lower bounds take the form of tail probabilities of $F$-distributions. They may be used to predict the threshold SNR. At a compression ratio of $C$, our numerical experiments show that the threshold SNR increases by about $10\mathrm{log}_{10}C~dB$ when estimating a broadside source DOA in interference located at half the Rayleigh limit of the pre-compressed array.

\bibliographystyle{IEEEtran}
\bibliography{ref_Journal}

\end{document}